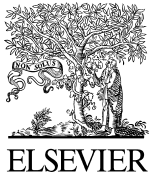

# A Water Mass Method and Its Application to Integrated Heat and Electricity Dispatch Considering Thermal Dynamics

Yuwei Chen[a], Qinglai Guo[a], Hongbin Sun[a], Zhengshuo Li[b], Zhaoguang Pan[a] and Wenchuan Wu[a]

[a]*Department of Electrical Engineering, State Key Laboratory of Electricity Systems, Tsinghua University, 100084 Beijing, China*
[b]*Shenzhen Environmental Science and New Energy Technology Engineering Laboratory, Tsinghua-Berkeley Shenzhen Institute (TBSI), Tsinghua University, Shenzhen, Guangdong, China*

**Abstract**

Currently, most district heating networks are running in a heat-setting mode, limiting the adjustment of the electrical power of combined heat and power (CHP) units. By considering the electrical power system (EPS) and district heating system (DHS) together, the peak regulatory capability of CHP units can be improved and renewable energy accommodation can be promoted. In this paper, a tractable integrated heat and electricity dispatch (IHED) model is described that addresses the thermal dynamic characteristics of pipelines and buildings to increase flexibility. To deal with the complexity of the optimization model, a water mass method (WMM) for pipeline thermal dynamics is proposed. Benefiting from the WMM, the proposed IHED model is an ordinary, non-linear model. An iterative algorithm based on the generalized Benders decomposition, and a sequential approach combined with the iterative algorithm and IPOPT, are proposed to solve the IHED model. Compared with a steady state model without thermal dynamics, considering the thermal dynamic characteristics in the DHS can further expand the peak regulatory capabilities of CHP units. The WMM is tested in the thermal dynamic simulations compared to an existing node method and a commercial simulation software. And the proposed solution strategy is verified in a small-scale system and a practical system. The simulation results of case studies are discussed to demonstrate the feasibility and economy of the dispatch model proposed here.

## 1. Introduction

With the rapid development of the global economy, the high dependence on 'traditional' energy makes the shortage of traditional fossil energy and environment problems increasingly serious [1]-[3]. Multi-energy systems (MES), also known as integrated energy systems (IES) or multi-carrier energy systems, have evolved rapidly in response to challenges of energy efficiency, carbon dioxide abatement, and renewable energy integration [4]-[9]. Multi-energy coupling can bring many benefits, but also make the original energy system more complex. District heating systems (DHSs) are common in northern China, where combined heat and power (CHP) units play an important role in heat supply. In Jilin Province, China, CHP units provide more the 70% of the heat load in winter. CHP units can produce heat and electricity at the same time, and they are the main coupling device between electrical power systems (EPSs) and DHSs. Generally, the operation of a CHP unit is determined by the heat load, and the flexibility to regulate power generation is thus limited. In the off-peak hours of winter, electricity demand is quite low whereas the heat load is high. To meet the heating demand, the CHP units must maintain and generate a certain amount of electricity, which results in high wind power curtailment (e.g., 89% of the total wind power curtailment in 2013 in Jilin Province was for this reason) [10]. Thus, increasing the operational flexibility of CHP units would improve wind energy use. Considering the thermal dynamics of a DHS is one way to decouple the electrical power and heat power of CHP units and to improve the dispatch flexibility of a EPS. There has been some research on integrated heat and electricity dispatch (IHED), including

---
Corresponding author at: Department of Electrical Engineering, Tsinghua University, Rm. 3-120, West Main Building, Tsinghua University, 100084 Beijing, China.
E-mail address: shb@tsinghua.edu.cn (H. Sun).

**Nomenclature**

*Indices, Sets*

| | |
|---|---|
| $b/\mathcal{B}$ | Index/set of pipelines. |
| $\mathcal{I}^{CHP}$ | Set of indices of CHP units. |
| $\mathcal{I}^{TU}$ | Set of indices of non-CHP units. |
| $\mathcal{I}^{RE}$ | Set of indices of renewable energy plants. |
| $l/\mathcal{L}$ | Index/set of lines in electrical power system. |
| $n/\mathcal{N}$ | Index/set of nodes in district heating system. |
| $n_p/\mathcal{N}_p$ | Index/set of buses in electrical power system. |
| $n_{1b}/n_{2b}$ | Index of node at the start/end of pipeline $b$. |
| $\mathcal{S}_n^{out}/\mathcal{S}_n^{in}$ | Set of indices of pipelines starting/ending at node $n$. |
| $\mathcal{S}_{n_p}^{CHP}/\mathcal{S}_{n_p}^{TU}$ | Set of CHP/non-CHP units located at bus $n_p$. |
| $\mathcal{S}_{n_p}^{RE}$ | Set of renewable energy plants located at bus $n_p$. |
| $\mathcal{T}$ | Set of indices of scheduling periods. |
| $\mathbb{Z}$ | Set of integers. |

*Parameters*

| | |
|---|---|
| $A_b$ | Cross-sectional area of pipeline $b$. |
| $a_{k,i}$ | $k$th cost coefficient of CHP unit $i$. |
| $c/c_{air}$ | Specific heat of water/air. |
| $D_{n_p}^\tau$ | Electric load demand at bus $n_p$ at period $\tau$. |
| $d_{k,i}$ | $k$th cost coefficient of non-CHP unit $i$. |
| $F_l$ | Transmission capacity limit of line $l$. |
| $G_a^\tau$ | Indoor ventilation at period $\tau$. |
| $H_k$ | Convective heat transfer coefficient between wall $k$ and indoor air. |
| $K_{l,n_p}$ | Shift factor of bus $n_p$ to line $l$. |
| $k_b$ | Resistance coefficient of pipeline $b$. |
| $k_{p,i}$ | Penalty factor of renewable energy curtailment of renewable energy plant $i$. |
| $L_b$ | Length of pipeline $b$. |
| $NK_i$ | Number of vertexes in feasible region of CHP unit $i$. |
| $N_b$ | Maximum number of periods it takes to flow through the pipeline $b$. |
| $N_s$ | Number of response factors. |
| $N_w$ | Number of walls. |
| $P_{k,i}/Q_{k,i}$ | Electricity/heat corresponding to the $k$th vertex of unit $i$. |
| $\Delta\overline{p}_i/\Delta\underline{p}_i$ | Upward/downward ramping capability of electricity output of unit $i$. |
| $\tilde{p}_{RE,i}^\tau$ | Available renewable energy power of plant $i$ at period $\tau$. |
| $\Delta\overline{q}_i/\Delta\underline{q}_i$ | Upward/downward ramping capability of heat output of CHP unit $i$. |
| $SR^u/SR^d$ | Required upward/downward spinning reserve capacity in system. |
| $S_k$ | Surface area of wall $k$. |
| $t_{am,b}^\tau$ | Ambient temperature of pipeline $b$ at period $\tau$. |
| $t_{out}^\tau$ | Outdoor temperature at period $\tau$. |
| $V$ | Volume of the room. |
| $\overline{X}/\underline{X}$ | Upper/Lower bound of variable $X$. |
| $Y_j, Z_j$ | Response factor. |
| $\Delta\tau$ | Time interval per period. |
| $\eta_{pump,b}$ | Efficiency of water pump. |
| $\lambda_b$ | Heat transfer coefficient of pipeline $b$. |
| $\rho/\rho_{air}$ | Density of water/air. |
| $\varphi_{i,k}^r$ | Radiation heat transfer conductivity between wall $k$ and wall $i$. |

*Functions and Variables*

| | |
|---|---|
| $C_i^{CHP}(\cdot)$ | Operation cost function of CHP unit $i$. |
| $C_i^{TU}(\cdot)$ | Operation cost function of non-CHP unit $i$. |
| $C_i^{RE}(\cdot)$ | Penalty cost of renewable energy plant $i$. |
| $h_n^\tau$ | Pressure at the node $n$ at period $\tau$. |
| $h_{pump,b}^\tau$ | Head of pump in the pipeline $b$ at period $\tau$. |
| $m_b^\tau$ | Mass flow rate of pipeline $b$ at period $\tau$. |
| $\boldsymbol{m}$ | Vector of mass flow rates. |
| $K_{b,\tau,k}$ | Coefficient variable in node method. |
| $p_{RE,i}^\tau$ | Generation of renewable energy plant $i$ at period $\tau$. |
| $p_i^\tau/q_i^\tau$ | Electricity/heat power output of unit $i$ at period $\tau$. |
| $q_n^\tau$ | Heat power injection at node $n$ at period $\tau$. |
| $q_{room}^\tau$ | Heat gain from indoor heat source at period $\tau$. |
| $ru_i^\tau/rd_i^\tau$ | Upward/downward spinning reserve capacity of non-CHP unit $i$ at period $\tau$. |
| $R_b^\tau, S_b^\tau$ | Auxiliary variables in node method. |
| $t_{s,b}^\tau/t_{e,b}^\tau$ | Temperatures at the start/end of pipeline $b$ at period $\tau$. |
| $t_{e',b}^\tau$ | Temperature at the end of pipeline $b$ without temperature drop. |
| $t_n^\tau$ | Temperature at node $n$ at period $\tau$. |
| $t_{wall,k}^\tau$ | Temperature of inner wall surface $k$ at period $\tau$. |
| $t_{room}^\tau$ | Temperature of indoor air at period $\tau$. |
| $\boldsymbol{x}$ | Vector of variables except $\boldsymbol{\alpha}_b^\tau, \boldsymbol{\beta}_b^\tau$ and $\boldsymbol{\eta}$. |
| $z_{b,k+1}^\tau$ | Binary variable in upper-level problem. |
| $\alpha_{b,k}^\tau, \beta_{b,k}^\tau$ | Weight coefficient of the water mass. |
| $\boldsymbol{\alpha}_b^\tau, \boldsymbol{\beta}_b^\tau$ | Vector of weight coefficient $\alpha_{b,k}^\tau, \beta_{b,k}^\tau$. |
| $\gamma_b^\tau, \phi_b^\tau$ | Auxiliary variables in node method. |
| $\zeta_{k,i}^\tau$ | $k$th combination factor of CHP $i$ at period $\tau$. |
| $\boldsymbol{\eta}$ | Vector of $\alpha_{b,k}^\tau$ and $\beta_{b,k}^\tau$. |
| $(\cdot)_r$ | Variables at iteration $r$. |

modeling and analyses. In most of these studies [11]-[20], the model of the heating network is simplified. Reference [11] proposed a model for centralized dispatch and explored the use of electrical boilers and heat storage tanks to increase the flexibility of CHP units. In [12], Yang et al. proposed a method to increase the flexibility of CHP units using thermal energy storage, distributed electric heat pumps, and buildings thermal inertia. However, references [11] and [12], as well as [12]-[15], do not consider a network model of a DHS. Reference [16] explored optimal ways to integrate electrical and heating systems to accommodate more renewable energy. Reference [17] formulates an integrated optimal power flow model of power and heat networks and presents an iterative algorithm to solve the model. Although references [16] and [17] consider a heating network model, the thermal dynamics of the DHS are ignored. References [18] and [19] studied the economic dispatch for integrated heat and electricity system considering the thermal dynamics of pipelines. In [20], based on a feasible region method, an IHED strategy considering building thermal dynamics was proposed. However, it was assumed in [18]-[20] that the mass flow rate was not involved in DHS regulation, which means that the flexibility of the DHS is not fully utilized. An optimization model considering the change in mass flow rate will be much more complicated and difficult to solve, so it is seldom adopted by the IHED. However, it has been proved by [21] that adjusting the mass flow can take advantage of a pipeline heat storage effect, and enhance dispatch flexibility to increase the wind power utilization ratio. In [21], a node method (NM) was used to account for the thermal dynamics in a water pipeline. Because some decision variables are involved in the NM as subscripts, the dispatch model is difficult to solve. The iterative solution strategy in [21] is heuristic and does not guarantee convergence for some cases.

In this paper, we focus on formulating a tractable IHED model considering the thermal dynamics and the adjustment of the mass flow, which can enhance the utilization of renewable energy. We make improvements in both models and algorithm, and present simulation results to show that our solution strategy has better performance than that in [21]. The main contribution of this paper can be summarized in the following points.

1) The NM model of pipeline used in [21] is complex and difficult to use for dispatch. Instead, we present a water mass method (WMM) for pipeline thermal dynamics with minimal influence on model accuracy. In the case studies, the WMM is tested in the thermal dynamic simulations compared to the existing NM and a commercial simulation software.

2) We propose an IHED model, considering pipeline and building thermal dynamics. To cope with this non-convex programming model and overcome the shortcomings of heuristic algorithm, a decomposition method and an efficient iterative algorithm are proposed based on a generalized Benders decomposition (GBD).

3) Because a large number of logical constraints such as $(1-\alpha_{b,k}^\tau)\alpha_{b,k+1}^\tau=0$ are included in the IHED model, the commercial solvers are difficult to directly solve the IHED model. Besides the iterative algorithm, we propose a sequential approach combined with our iterative algorithm and IPOPT [22] to improve the optimality of the solution.

The remainder of this paper is organized as follows. Section 2 presents a model of the DHS, including the models of source, load and network. In Section 3, a WMM for pipeline thermal dynamics is proposed and we explain the physical meaning of this method. In Section 4, the IHED model incorporating the pipeline and building thermal dynamics model is established. Section 5 describes the decomposition of the IHED model and describes an iterative algorithm. In Section 6, case studies are presented to show the effectiveness of the proposed model and solution strategy. Finally, Section 7 concludes the research and discusses future work.

## 2. District Heating System Model

### 2.1. Heating source

The feasible operating regions of a CHP unit can be described using convex polygons. The electricity power $p_i^\tau$ and heat power $q_i^\tau$ of CHP unit are represented using a combination of vertexes in the polygonal feasible operating region [23],

$$p_i^\tau = \sum_{k=1}^{NK_i} \zeta_{k,i}^\tau P_{k,i}, q_i^\tau = \sum_{k=1}^{NK_i} \zeta_{k,i}^\tau Q_{k,i}, \forall i \in \mathcal{I}^{CHP}, \tau \in \mathcal{T} \quad (1)$$

where combination factor $\zeta_k$ satisfies the following formula,

$$\sum_{k=1}^{NK_i} \zeta_{k,i}^\tau = 1, 0 \leq \zeta_{k,i}^\tau \leq 1, \\ \forall k \in \{1,2,...,NK_i\}, i \in \mathcal{I}^{CHP}, \tau \in \mathcal{T} \quad (2)$$

The operation cost of a CHP unit is expressed as a quadratic function of the electricity and heat powers [23]:

$$C_i^{CHP}(p_i^\tau, q_i^\tau) = a_{0,i} + a_{1,i} p_i^\tau + a_{2,i} q_i^\tau + a_{3,i}(p_i^\tau)^2 \\ + a_{4,i}(q_i^\tau)^2 + a_{5,i} p_i^\tau q_i^\tau, \forall i \in \mathcal{I}^{CHP}, \tau \in \mathcal{T} \quad (3)$$

Function (3) is convex when $4a_{3,i}a_{4,i} \geq a_{5,i}^2$. Generally, this inequality holds for CHP units [23]-[26]. Additionally, the ramping constraints of CHP units are considered:

$$-\Delta \underline{p}_i \Delta \tau \leq p_i^\tau - p_i^{\tau-1} \leq \Delta \overline{p}_i \Delta \tau, \forall i \in \mathcal{I}^{CHP}, \tau \in \mathcal{T}, \quad (4)$$

$$-\Delta \underline{q}_i \Delta \tau \leq q_i^\tau - q_i^{\tau-1} \leq \Delta \overline{q}_i \Delta \tau, \forall i \in \mathcal{I}^{CHP}, \tau \in \mathcal{T}. \quad (5)$$

### 2.2. Heating load

The response factor method [27] is used to establish the heat balance model of the building. This is a widely used and accurate simulation model for building thermal dynamics.

The heat balance equation for wall $i$ is given below:

$$\sum_{j=0}^{N_s} Y_j t_{out}^{\tau-j} - \sum_{j=0}^{N_s} Z_j t_{wall,i}^{\tau-j} + \sum_{k=1}^{N_i} \varphi_{i,k}^r \left(t_{wall,k}^\tau - t_{wall,i}^\tau\right) \\ + H_i \left(t_{room}^\tau - t_{wall,i}^\tau\right) = 0 \quad (6)$$

where $\sum_{j=0}^{N_s} Y_j t_{out}^{\tau-j} - \sum_{j=0}^{N_s} Z_j t_{wall,i}^{\tau-j}$, $\sum_{k=1}^{N_i} \varphi_{i,k}^r \left(t_{wall,k}^\tau - t_{wall,i}^\tau\right)$, and $H_i\left(t_{room}^\tau - t_{wall,i}^\tau\right)$ denote heat transfer via conduction, radiation, and convection, respectively.

The indoor air heat balance equation is as follows:

$$\sum_{k=1}^{N_w} S_k H_k \left(t_{wall,k}^\tau - t_{room}^\tau\right) - G_a^\tau c_{air} \rho_{air} \left(t_{room,i}^\tau - t_{out}^\tau\right) \\ + q_{room}^\tau = V c_{air} \rho_{air} \left(t_{room}^\tau - t_{room}^{\tau-1}\right) \quad (7)$$

This equation indicates that "Heat transfer via conduction" – "Heat loss by fresh wind supply" + "Heat gain from indoor heat source" = "Incremental heat of indoor air".

In order to maintain indoor thermal comfort, the temperature of indoor air should not exceed the bound.

$$\underline{t}_{room}^\tau \leq t_{room}^\tau \leq \overline{t}_{room}^\tau \quad (8)$$

### 2.3. Heating network

1) Pressure difference: The pressure difference between outset $n_{1b}$ and endpoint $n_{2b}$ of pipeline is expressed as

$$h_{n_{1b,b}}^\tau - h_{n_{2b,b}}^\tau = k_b (m_b^\tau)^2 - h_{pump,b}^\tau, \forall b \in \mathcal{B}, \tau \in \mathcal{T}. \quad (9)$$

$$\underline{h}_{pump,b}^\tau \leq h_{pump,b}^\tau \leq \overline{h}_{pump,b}^\tau, \forall b \in \mathcal{B}, \tau \in \mathcal{T}. \quad (10)$$

In (9), we assume that the mass flow direction is determined. If there is no water pump in the pipeline, $h_{pump,b}^\tau$ equals zero.

Additionally, the power consumed by a water pump is as follows:

$$p_{pump,b}^{\tau} = \frac{m_b^{\tau} h_{pump,b}^{\tau}}{\rho \eta_{pump,b}}, \forall b \in \mathcal{B}, \tau \in \mathcal{T}. \quad (11)$$

2) Continuity of mass flow: Assuming that water is incompressible and no leakage or consumption of water occurs, the total mass flow that comes into a node is equal to the total mass flow that leaves the node.

$$\sum_{b \in \mathcal{S}_n^{out}} m_b^{\tau} = \sum_{b \in \mathcal{S}_n^{in}} m_b^{\tau}, \forall n \in \mathcal{N}, \tau \in \mathcal{T}. \quad (12)$$

3) Mass flow rate limit: Mass flow rates in the pipeline should not exceed the bound.

$$\underline{m}_b \leq m_b^{\tau} \leq \overline{m}_b, \forall b \in \mathcal{B}, \tau \in \mathcal{T}. \quad (13)$$

4) Temperature mixing: We assume that the mass flow enters into the same node, is mixed, and the temperature at the start of each pipeline leaving the node is equal to the mixture temperature at the node. The relationship between the temperature at the node and the temperature of each pipeline is expressed as follows:

$$c \sum_{b \in \mathcal{S}_n^{out}} m_b^{\tau} t_n^{\tau} = c \sum_{b \in \mathcal{S}_n^{in}} m_b^{\tau} t_{e,b}^{\tau} + q_n^{\tau}, \forall n \in \mathcal{N}, \tau \in \mathcal{T}, \quad (14)$$

$$t_{s,b}^{\tau} = t_n^{\tau}, \forall b \in \mathcal{S}_n^{out}, \tau \in \mathcal{T}, \quad (15)$$

where $q_n^{\tau}$ is the heat power injection at node $n$ at period $\tau$.

5) Thermal dynamics and heat loss: In the water pipeline, changes in inlet temperature spread to the outlet slowly, which covers approximately the time of water flow through the pipeline [28]. Additionally, due to the temperature difference between the flowing water and its surrounding environment, the temperature of the water is reduced. The model of temperature changes along the pipeline is the most complex part of the DHS model, which is the core challenge that has to be addressed if we want to consider the thermal dynamics into the IHED. So, it will be described in the next section in detail.

## 3. Water Mass Method

In [21], Li et al. used the NM to account for the thermal dynamics and heat loss of the water pipeline. The NM (16)-(21) is an accurate method for temperature dynamic simulations [28]-[29].

$$\gamma_b^{\tau} = \min_{\gamma} \left\{ \gamma : s.t. \sum_{k=0}^{\gamma} m_b^{\tau-k} \Delta \tau \geq \rho A_b L_b, \gamma \geq 0, \gamma \in \mathbb{Z} \right\}, (16)$$

$$\phi_b^{\tau} = \min_{\phi} \left\{ \phi : s.t. \sum_{k=1}^{\phi} m_b^{\tau-k} \Delta \tau \geq \rho A_b L_b, \phi \geq 0, \phi \in \mathbb{Z} \right\}, (17)$$

$$R_b^{\tau} = \sum_{k=0}^{\gamma_b^{\tau}} \left( m_b^{\tau-k} \Delta t \right), S_b^{\tau} = \begin{cases} \sum_{k=0}^{\phi_b^{\tau}-1} \left( m_b^{\tau-k} \Delta \tau \right), \text{if } \phi_b^{\tau} \geq \gamma_b^{\tau}+1 \\ R_b^{\tau}, \text{otherwise} \end{cases}, (18)$$

$$K_{b,\tau,k} = \begin{cases} (m_b^{\tau}\Delta\tau + \rho A_b L_b - S_b^{\tau})/m_b^{\tau}\Delta\tau, k = \tau-\phi_b^{\tau} \\ m_b^k/m_b^{\tau}, k = \tau-\phi_b^{\tau}+1,...,\tau-\gamma_b^{\tau}-1 \\ (R_b^{\tau} - \rho A_b L_b)/m_b^{\tau}\Delta\tau, k = \tau-\gamma_b^{\tau} \\ 0, \text{otherwise} \end{cases}, (19)$$

$$t_{e',b}^{\tau} = \sum_{k=\tau-\phi_b^{\tau}}^{\tau-\gamma_b^{\tau}} K_{b,\tau,k} t_{s,b}^k, \quad (20)$$

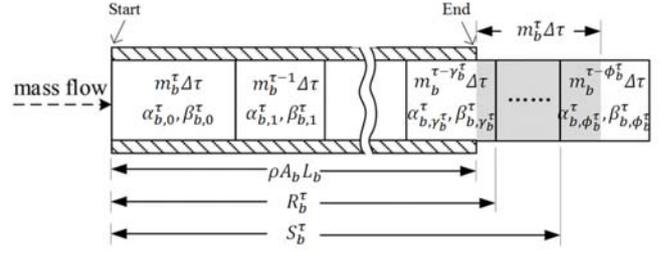

Fig. 1. Vertical section of a pipeline [21]

$$t_{e,b}^{\tau} = t_{am,b}^{\tau} + (t_{e',b}^{\tau} - t_{am,b}^{\tau})$$
$$\exp\left[-\frac{\lambda_b \Delta \tau}{A_b \rho c}\left(\gamma_b^{\tau} + \frac{1}{2} + \frac{S_b^{\tau} - R_b^{\tau}}{m_b^{\tau-\gamma_b^{\tau}} \Delta \tau}\right)\right]. \quad (21)$$

The NM was designed originally for heating system simulations. However, it is difficult to use (16)–(21) in optimization models, because the integer variables $\gamma_b^{\tau}$ and $\phi_b^{\tau}$ are solved by optimization problems (16)-(17) and are involved in (18)-(21) as subscripts. Here, we seek to provide an equivalent model, referred to as the WMM.

$$\sum_{k=0}^{N_b} \alpha_{b,k}^{\tau} m_b^{\tau-k} \Delta \tau = \rho A_b L_b, \quad (22)$$

$$(1-\alpha_{b,k}^{\tau})\alpha_{b,k+1}^{\tau} = 0, k = 0,1,...,N_b-1, \quad (23)$$

$$0 \leq \alpha_{b,k}^{\tau} \leq 1, k = 0,1,...,N_b, \quad (24)$$

$$\sum_{k=1}^{N_b} \beta_{b,k}^{\tau} m_b^{\tau-k} \Delta \tau = \rho A_b L_b, \quad (25)$$

$$(1-\beta_{b,k}^{\tau})\beta_{b,k+1}^{\tau} = 0, k = 1,2,...,N_b-1, \quad (26)$$

$$\beta_{b,0}^{\tau} = 1, \ 0 \leq \beta_{b,k}^{\tau} \leq 1, k = 1,1,...,N_b, \quad (27)$$

$$m_b^{\tau} t_{e',b}^{\tau} = \sum_{k=0}^{N_b} (\beta_{b,k}^{\tau} - \alpha_{b,k}^{\tau}) m_b^{\tau-k} t_{s,b}^{\tau-k}, \quad (28)$$

$$t_{e,b}^{\tau} = t_{am,b}^{\tau} + (t_{e',b}^{\tau} - t_{am,b}^{\tau})$$
$$\exp\left[-\frac{\lambda_b \Delta \tau}{2A_b \rho c}\left(\sum_{k=0}^{N_b}\alpha_{b,k}^{\tau} + \sum_{k=1}^{N_b}\beta_{b,k}^{\tau}\right)\right]. \quad (29)$$

This WMM model (22)-(29) eliminates 'min' and integer variables that occur in the subscripts. Thus, it can be used more readily in optimization models.

Below, we give a brief explanation of the physical meaning of NM (16)-(21) and WMM (22)-(29). In Fig. 1, $m_b^{\tau}\Delta\tau$ indicates the mass of water flowing into the pipeline in period $\tau$. The main concepts of NM and WMM include the following two steps.

The first is to estimate the lossless outlet temperature, $t_{e',b}^{\tau}$, by averaging the temperature of the water mass flowing out of the pipeline during period $\tau$, which is shown graphically by the gray part in Fig. 1. The vectors $\boldsymbol{\alpha}_b^{\tau}$ and $\boldsymbol{\beta}_b^{\tau}$ of WMM are the weight coefficients of the water mass. Each element of $\boldsymbol{\alpha}_b^{\tau}$ and $\boldsymbol{\beta}_b^{\tau}$ belong to interval $[0,1]$. Constraints (23)-(24) limit the vector $\boldsymbol{\alpha}_b^{\tau}$ to a sequence, like $(1,...1,x,0,0...)$, where $x \in [0,1]$. Similarly, the vector $\boldsymbol{\beta}_b^{\tau}$ has the same characteristics. $\boldsymbol{\alpha}_b^{\tau}$ represents the weight of the water mass filled in the pipeline. $\boldsymbol{\beta}_b^{\tau}$ represents the weight of the water mass filled in the pipeline and the gray part. Thus, the vector $\boldsymbol{\beta}_b^{\tau} - \boldsymbol{\alpha}_b^{\tau}$ represents the weight of the water mass constituting the gray part, and formulae (22)-(28) are strictly equivalent to (16)–(20).

Then, $t_{e',b}^{\tau}$ is reduced to reflect the heat loss for the pipeline to the surroundings. In (21), $(\gamma_b^{\tau}\Delta\tau+\Delta\tau/2+(S_b^{\tau}-R_b^{\tau})/m_b^{\tau-\gamma_b^{\tau}})$ is an estimate of the time it takes to flow through the pipeline. In (29), we use another expression, $\left(\sum_{k=0}^{N_b}\alpha_{b,k}^{\tau}+\sum_{k=1}^{N_b}\beta_{b,k}^{\tau}\right)\Delta\tau/2$, to estimate this time. These two expressions tend to be equal as $\Delta\tau$ small enough.

In order to better explain the WMM and illustrate the equivalence between formulae (22)-(28) and (16)–(20), we present an example of a single pipeline below.

Assuming $L_b=1750\text{m}$, $A_b=0.5\text{m}^2$, $t_{am,b}^{\tau}=10°\text{C}$, $\Delta\tau=1\text{h}$ and $\tau=12$. Physical parameters of the pipeline are as follows: $\rho=10^3\text{kg/m}^3$, $c=4.2\times10^3\text{J/(kg·K)}$ and $\lambda_b=0.12\text{W/(m·K)}$. The historical temperature and mass flow data are shown in Table 1. The temperatures at the start of pipeline in Table 1 are designed to be make the results more distinguishable.

Table 1. Historical temperatures and mas flow rates

| Mass flow | Value (kg/s) | Temperatures at the start of pipeline | Value (°C) |
|---|---|---|---|
| $m_b^9$ | 116.10 | $t_{s,b}^9$ | 80 |
| $m_b^{10}$ | 113.68 | $t_{s,b}^{10}$ | 90 |
| $m_b^{11}$ | 185.52 | $t_{s,b}^{11}$ | 100 |
| $m_b^{12}$ | 120.21 | $t_{s,b}^{12}$ | 110 |

Both the NM and WMM are used to calculate $t_{e,b}^{12}$ that is the temperature at the end of pipeline $b$ at period 12. The coefficients $K_{b,12,k}(k\geq9)$ and temperatures $t_{e',b}^{12},t_{e,b}^{12}$ of NM are shown in Table 2. The weight coefficients $\alpha_{b,k}^{12}(k\leq3)$, $\beta_{b,k}^{12}(k\leq3)$ and temperature $t_{e',b}^{12},t_{e,b}^{12}$ of WMM are shown in Table 3. $t_{e',b}^{12}$ of NM and of WMM are calculated by (20) and (28), respectively. $t_{e,b}^{12}$ of NM and that of WMM are calculated by (21) and (29), respectively. The estimated time of NM it takes to flow through the pipeline and that of WMM are also shown in Table 2 and Table 3, respectively. The coefficients $K_{b,12,k}(k<9)$, $\alpha_{b,k}^{\tau}(k>3)$ and $\beta_{b,k}^{12}(k>3)$ not listed in tables below are all equal to 0.

As shown in Table 2 and Table 3, $t_{e',b}^{12}$ of NM and that of WMM are strictly equal; $t_{e,b}^{12}$ of NM and that of WMM are almost equal. Additionally, it will be shown in the case studies that the estimated time of WMM may be better than that of NM.

Table 2. Results of NM

| Coefficients | Value | Temperatures | Value (°C) |
|---|---|---|---|
| $K_{b,\tau,9}$ | 0 | $t_{e',b}^{12}$ | 95.219 |
| $K_{b,\tau,10}$ | 0.4786 | $t_{e,b}^{12}$ | 95.193 |
| $K_{b,\tau,11}$ | 0.5214 | Estimated time | Value (h) |
| $K_{b,\tau,12}$ | 0 | $\left(\gamma_b^{\tau}+\frac{1}{2}+\frac{S_b^{\tau}-R_b^{\tau}}{m_b^{\tau-\gamma_b^{\tau}}\Delta\tau}\right)\Delta\tau$ | 1.5 |

Table 3. Results of WMM

| Coefficients | Value | Coefficients | Value | Temperatures | Value (°C) |
|---|---|---|---|---|---|
| $\alpha_{b,0}^{12}$ | 1 | $\beta_{b,0}^{12}$ | 1 | $t_{e',b}^{12}$ | 95.219 |
| $\alpha_{b,1}^{12}$ | 0.6621 | $\beta_{b,1}^{12}$ | 1 | $t_{e,b}^{12}$ | 95.194 |
| $\alpha_{b,2}^{12}$ | 0 | $\beta_{b,2}^{12}$ | 0.5061 | Estimated time | Value (h) |
| $\alpha_{b,3}^{12}$ | 0 | $\beta_{b,3}^{12}$ | 0 | $\frac{\Delta\tau}{2}\left(\sum_{k=0}^{N_b}\alpha_{b,k}^{\tau}+\sum_{k=1}^{N_b}\beta_{b,k}^{\tau}\right)$ | 1.42 |

## 4. Formulation of IHED Model

### 4.1. Objective Function

The objective function (30) minimizes the total operational cost over 24 h. The cost contains the operating cost of CHP units (3) and non-CHP units and the renewable energy penalty costs.

$$\min\sum_{\tau\in\mathcal{T}}\sum_{i\in\mathcal{I}^{CHP}}C_i^{CHP}+\sum_{\tau\in\mathcal{T}}\sum_{i\in\mathcal{I}^{TU}}C_i^{TU}+\sum_{\tau\in\mathcal{T}}\sum_{i\in\mathcal{I}^{RE}}C_i^{RE}. \quad(30)$$

$C_i^{TU}$ is a quadratic function of power output:

$$C_i^{TU}(p_i^{\tau})=d_{0,i}+d_{1,i}p_i^{\tau}+d_{2,i}(p_i^{\tau})^2,\forall i\in\mathcal{I}^{TU},\tau\in\mathcal{T}. \quad(31)$$

$C_i^{RE}$ is the penalty term for renewable energy curtailment:

$$C_i^{RE}(p_{RE,i}^{\tau})=k_{p,i}(\tilde{p}_{RE,i}^{\tau}-p_{RE,i}^{\tau})^2,\forall i\in\mathcal{I}^{CHP},\tau\in\mathcal{T}. \quad(32)$$

### 4.2. Constraints

1) Power flow constraints: We use the DC power flow model to establish a power flow equation:

$$-F_l\leq\sum_{n_p\in\mathcal{N}_p}K_{l,n_p}(\sum_{i\in S_{n_p}^{CHP}\cup S_{n_p}^{TU}}p_i^{\tau}+\sum_{i\in S_{n_p}^{RE}}p_{RE,i}^{\tau}-D_{n_p}^{\tau})\leq F_l,$$
$$\forall l\in\mathcal{L},\tau\in\mathcal{T} \quad(33)$$

2) Generation output constraints of renewable energy plants: The power output of a renewable energy plant is limited by its available renewable energy power:

$$0\leq p_{RE,i}^{\tau}\leq\tilde{p}_{RE,i}^{\tau},\forall i\in\mathcal{I}^{CHP},\tau\in\mathcal{T}. \quad(34)$$

3) CHP unit operational constraints: Defined in (1) and (2).
4) CHP unit ramping constraints: Defined in (5).
5) Generation output constraints of non-CHP units: Power outputs of non-CHP units are within their technical limits,

$$\underline{p}_i\leq p_i^{\tau}\leq\overline{p}_i,\forall i\in\mathcal{I}^{TU},\tau\in\mathcal{T}. \quad(35)$$

6) Non-CHP unit ramping constraints: The form of expression is analogous to (4) and (5).

$$-\Delta\underline{p}_i\Delta\tau\leq p_i^{\tau}-p_i^{\tau-1}\leq\Delta\overline{p}_i\Delta\tau,\forall i\in\mathcal{I}^{TU},\tau\in\mathcal{T}. \quad(36)$$

7) Spinning reserve constraints: Non-CHP units need to provide sufficient spinning reserve capacity

$$0\leq ru_i^{\tau}\leq\Delta\overline{p}_i\Delta\tau,ru_i^{\tau}\leq\overline{p}_i-p_i^{\tau},\forall i\in\mathcal{I}^{TU},\tau\in\mathcal{T}, \quad(37)$$

$$0\leq rd_i^{\tau}\leq\Delta\underline{p}_i\Delta\tau,rd_i^{\tau}\leq p_i^{\tau}-\underline{p}_i,\forall i\in\mathcal{I}^{TU},\tau\in\mathcal{T}, \quad(38)$$

$$\sum_{i\in\mathcal{I}^{TU}}ru_i^{\tau}\geq SR^{up},\sum_{i\in\mathcal{I}^{TU}}rd_i^{\tau}\geq SR^{down},\forall\tau\in\mathcal{T}. \quad(39)$$

8) DHS operating constraints: Defined in (6)-(15) and (22)-(29).

9) DHS security constraints: The node temperature and node pressure should not exceed their upper and lower bounds

$$\underline{t}_n^{\tau}\leq t_n^{\tau}\leq\overline{t}_i^{\tau},\forall n\in\mathcal{N},\tau\in\mathcal{T}, \quad(40)$$

$$\underline{h}_n^{\tau}\leq h_n^{\tau}\leq\overline{h}_n^{\tau},\forall n\in\mathcal{N},\tau\in\mathcal{T}. \quad(41)$$

## 5. Solution Strategy

Applying the WMM, the IHED model described in Section 4 is a nonlinear programming problem and sometimes can be solved directly by IPOPT. However, because a large number of logical constraints such as (23) are included in the IHED model, it is difficult to solve the IHED model by IPOPT in many cases. In order to solve the IHED effectively, we propose a decomposition method and an iterative algorithm based on the GBD. Besides the iterative algorithm, a sequential approach can be used to improve the optimality of the result.

### 5.1. Model Decomposition

The vector of $\alpha_{b,k}^\tau$ and $\beta_{b,k}^\tau$ is denoted by $\eta$. $m$ is the mass flow vector. The vector of other variables is denoted by $x$. $f_2$ and $g_2$ represent the constraints of WMM (22)-(28), where only contain the variables $\eta$ and $m$. The IHED model can be expressed in the following form:

$$\begin{aligned} \min \ & c(x) \\ s.t. \ & g_1(x,m,\eta) = 0 \\ & f_2(m,\eta) \leq 0, g_2(m,\eta) = 0 \\ & x \in D_x, m \in D_m \end{aligned} \quad (42)$$

where $c(x)$ is a convex function. $D_x$ and $D_m$ are convex sets. Constraints $g_1$, $f_2$ and $g_2$ have the following characteristics:

1) $g_1$ is linear in $x$ for fixed $m$ and $\eta$.
2) For fixed $m$, the set $D = \{\eta \mid f_2(m,\eta) \leq 0, g_2(m,\eta) = 0\}$ contains a unique element. Thus, $\eta$ can be regarded as a mapping from $m$, denoted by $\eta(m)$.

For model decomposition, we regard $m$ and $\eta$ as complex variables and apply the GBD [30]. The subscript $r$ is used to denote the iterative number. The subproblem (SP) and master problem (MP) are as follows:

$$\text{SP:} \min_{x \in D_x} c(x) \\ s.t. \ g_1(x,m_r,\eta_r) = 0 \quad (43)$$

$$\text{MP:} \min_{m \in D_m, \eta, \mu_B} \mu_B \\ s.t. \ \mu_B \geq \xi_j(m,\eta;\mu_j), j \in I_p \\ 0 \geq \bar{\xi}_j(m,\eta;\bar{\mu}_j), j \in I_q \\ f_2(m,\eta) \leq 0, g_2(m,\eta) = 0 \quad (44)$$

where $\xi_j$ and $\bar{\xi}_j$ are the support functions of GBD.

If the SP is feasible, we can formulate the Lagrange function (45) and support function (46):

$$L(x,m,\eta;\mu_r) = \mu_r^T g_1(x,m,\eta) + c(x), \quad (45)$$

$$\xi_r(m,\eta;\mu_r) = \inf_{x \in D_x} L(x,m,\eta;\mu_r). \quad (46)$$

If the SP is infeasible, we solve the feasibility problem (FP).

$$\text{FP:} \min_{x \in D_x, a^+, a^-} 1^T a^+ + 1^T a^- \\ s.t. \ g_{1,i}(x,m_r,\eta_r) = 0, i \notin I \\ g_{1,i}(x,m_r,\eta_r) = a_i^+ - a_i^-, i \in I \\ a^+ \geq 0, a^- \geq 0 \quad (47)$$

where $I$ is the set of relaxed constraints. A viable option for $I$ is the set of temperature mixing constraints (14). The Lagrange function and support function resulting from the feasibility problem are defined by (48) and (49), respectively.

$$\bar{L}(x,m,\eta;\bar{\mu}_r) = \bar{\mu}_r^T g_1(x,m,\eta), \quad (48)$$

$$\bar{\xi}_r(m,\eta;\bar{\mu}_r) = \inf_{x \in D_x} \bar{L}(x,m,\eta;\bar{\mu}_r). \quad (49)$$

Reference [31] presents a relaxed master problem of GBD combined with the idea of an outer approximation algorithm. According to [31], we used a first-order Taylor polynomial of $L(x,m,\eta;\mu_r)$ and $\bar{L}(x,m,\eta;\bar{\mu}_r)$ around the point $(x_r,m_r,\eta_r)$ as $\xi_r(m,\eta;\mu_r)$ and $\bar{\xi}_r(m,\eta;\bar{\mu}_r)$.

$$\xi_r(m,\eta;\mu_r) = \mu_r^T \left( \frac{\partial g_1}{\partial m}(x_r,m_r,\eta_r)(m-m_r) \right. \\ \left. + \frac{\partial g_1}{\partial \eta}(x_r,m_r,\eta_r)(\eta-\eta_r) + g_1(x_r,m_r,\eta_r) \right) + c(x_r) \quad (50)$$

$$\bar{\xi}_r(m,\eta;\bar{\mu}_r) = \bar{\mu}_r^T \left( \frac{\partial g_1}{\partial m}(x_r,m_r,\eta_r)(m-m_r) \right. \\ \left. + \frac{\partial g_1}{\partial \eta}(x_r,m_r,\eta_r)(\eta-\eta_r) + g_1(x_r,m_r,\eta_r) \right) \quad (51)$$

Because $\eta$ can be regarded as a mapping from $m$ denoted by $\eta(m)$, $\xi_r(m,\eta;\mu_r)$ and $\bar{\xi}_r(m,\eta;\bar{\mu}_r)$ can be further relaxed as

$$\xi_r(m;\mu_r) = \mu_r^T g_1(x_r,m_r,\eta_r) + \mu_r^T \left( \frac{\partial g_1}{\partial m}(x_r,m_r,\eta_r) \right. \\ \left. + \frac{\partial g_1}{\partial \eta}(x_r,m_r,\eta_r) \frac{\partial \eta}{\partial m}(m_r) \right)(m-m_r) + c(x_r) \quad (52)$$

$$\bar{\xi}_r(m;\bar{\mu}_r) = \bar{\mu}_r^T g_1(x_r,m_r,\eta_r) + \bar{\mu}_r^T \left( \frac{\partial g_1}{\partial m}(x_r,m_r,\eta_r) \right. \\ \left. + \frac{\partial g_1}{\partial \eta}(x_r,m_r,\eta_r) \frac{\partial \eta}{\partial m}(m_r) \right)(m-m_r) \quad (53)$$

where $\partial \eta / \partial m$ can be obtained directly from its physical meaning. The non-zero elements of $\partial \eta / \partial m$ are

$$\frac{\partial \alpha_{b,k}^\tau}{\partial m_b^{\tau-j}} = \frac{1}{m_b^{\tau-k}}, if\ 0 < \alpha_{b,k}^\tau < 1\ and\ j \leq k, \quad (54)$$

$$\frac{\partial \beta_{b,k}^\tau}{\partial m_b^{\tau-j}} = \frac{1}{m_b^{\tau-k}}, if\ 0 < \beta_{b,k}^\tau < 1\ and\ j \leq k. \quad (55)$$

The discontinuities of the partial derivatives are ignored in (54) and (55), when $(\alpha_{b,k-1}^\tau = 1, \alpha_{b,k}^\tau = 0)$ or $(\beta_{b,k-1}^\tau = 1, \beta_{b,k}^\tau = 0)$.

Equations (52) and (53) are only related to $m$. Recall that $D = \{\eta \mid f_2(m_r,\eta) \leq 0, g_2(m_r,\eta) = 0\}$ contains a unique element. As such, the MP can be calculated in two steps. The first is to obtain $m_r$ and $\mu_B$ by solving the lower-level problem (LLP). Then, for fixed $m_r$, we solve the upper-level problem (ULP) to obtain the $\eta_r$. The LLP and the ULP of the MP are as follows:

$$\text{LLP:} \min_{m \in D_m, \mu_B} \mu_B \\ s.t. \ \mu_B \geq \xi_j(m;\mu_j), j \in I_p, \\ 0 \geq \bar{\xi}_j(m;\bar{\mu}_j), j \in I_q \quad (56)$$

$$\text{ULP:} \min_{\boldsymbol{\eta}} \quad 0$$
$$\text{s.t.} \quad f_2(\boldsymbol{m}_r, \boldsymbol{\eta}) \leq 0 \ . \tag{57}$$
$$g_2(\boldsymbol{m}_r, \boldsymbol{\eta}) = 0$$

The non-linear constraints in the ULP are (23) and (26). For example, constraints (23) and (24) can be reformulated as follows by adding binary variables:

$$0 \leq \alpha_{b,k+1}^{\tau} \leq z_{b,k+1}^{\tau}, k = 0,1,...,N_b - 1, \tag{58}$$
$$0 \leq 1 - \alpha_{b,k}^{\tau} \leq 1 - z_{b,k+1}^{\tau}, k = 0,1,...,N_b - 1, \tag{59}$$

where $z_{b,k+1}^{\tau}$ is a binary variable.

According to these transformations, the SP, FP and LLP are linear constrained programming problems, and the ULP is a mixed-integer linear programming (MILP) problem.

*5.2. Algorithm*

Based on the discussion above, an iterative algorithm is given below.

Step 1. Initialize a mass flow vector $\boldsymbol{m}_1$. Solve the ULP and determine the vector $\boldsymbol{\eta}_1$. Set the counter $r = 1$. Set the sets $I_p = \varnothing$ and $I_q = \varnothing$. Initialize the upper bound *UBD* as positive infinity, the lower bound *LBD* as zero. Select the convergence tolerance $\varepsilon > 0$.

Step 2. For fixed $\boldsymbol{m}_r$ and $\boldsymbol{\eta}_r$, solve the SP.

Step 2a. Feasible SP.

The SP has the objective value $v(\boldsymbol{m}_r, \boldsymbol{\eta}_r)$, which is finite with an optimal solution $\boldsymbol{x}_r$ and an optimal multiplier vector $\boldsymbol{\mu}_r$. Update $UBD = \min\{UBD, v(\boldsymbol{m}_r, \boldsymbol{\eta}_r)\}$. If $(UBD - LBD) / UBD < \varepsilon$, then terminate. Otherwise, generate the support function $\xi_r$ and add $r$ to the set $I_p$. Update $r = r + 1$.

Step 2b. Infeasible SP.

The SP is infeasible. Solve the FP to obtain the optimal solution $\boldsymbol{x}_r$ and the multiplier vector $\bar{\boldsymbol{\mu}}_r$. Generate the support function $\bar{\xi}_r$, and add $r$ to the set $I_q$. Update $r = r + 1$.

Step 3. Solve the LLP. Update the lower bound *LBD*, as the objective value of the LLP. Create variable $\boldsymbol{m}_r$ with the solution of the LLP.

Step 4. Solve the ULP for fixed $\boldsymbol{m}_r$. Create variable $\boldsymbol{\eta}_r$ with the solution of the ULP and return to Step 2.

*5.3. Sequential Approach*

Because the support functions in the MP are relaxed into (52) and (53), there is no guarantee that the solution of the above algorithm is globally optimal or locally optimal. However, a sub-optimal solution can be obtained by the iterative algorithm and the decision makers may believe that this sub-optimal solution is not too "far away" from the global optimum or local optimum. Below, we present a sequential approach combined with the iterative algorithm and IPOPT to improve the optimality of the solution. This sequential approach contains following two steps.

The first is to obtain a sub-optimal solution $(\boldsymbol{x}_r, \boldsymbol{m}_r, \boldsymbol{\eta}_r)$ by using the iterative algorithm. Then, we switch to use IPOPT, which implements a primal-dual interior point method. IPOPT is used to obtain a local optimum with initial point $(\boldsymbol{x}_r, \boldsymbol{m}_r, \boldsymbol{\eta}_r)$.

**6. Case Studies**

In this section, case studies to test the proposed WMM and the iterative algorithm are described. All tests are performed on a computer with a quad-core processor running at 2.40 GHz with 4 GB of memory. Programs are coded using MATLAB R2013b. All SPs, FPs, LLPs and ULPs are solved with CPLEX [32].

*6.1. Thermal Dynamic Simulations*

We conduct thermal dynamic simulations on a single pipeline using the WMM, the NM and Bentley sisHYD [33] that is a commercial software. The parameters of pipeline are in [34]. The mass flow rates and the temperatures at the start/end of pipeline, which are taken from the simulation results of Bentley sisHYD, can also be found in [34]. Fig. 2 shows the mass flow rates and the temperatures at the start of the pipeline. Assuming the data in Fig. 2 are known, the WMM and the NM are both used for calculating the temperatures at the end of pipeline. The time step size is 3 minutes and the results are shown in Fig. 3.

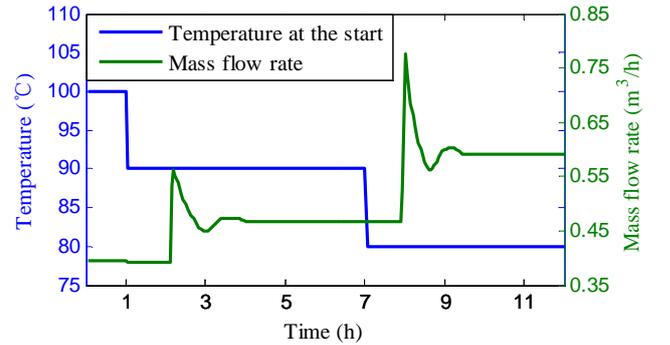

Fig. 2. Mass flow rates and temperatures at the start of pipeline.

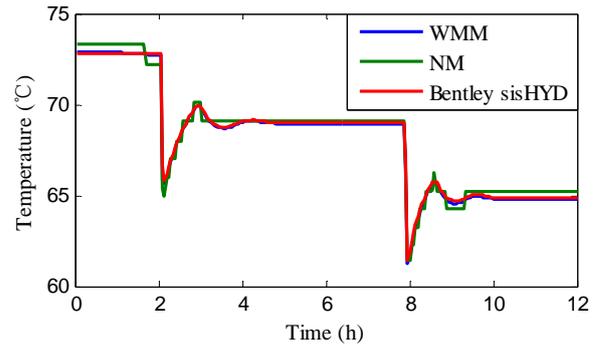

Fig. 3. Temperatures at the end of pipeline.

As shown in Fig. 3, the curves of the temperatures at the end of pipeline calculated by the WMM and the NM are both close to the curve calculated by Bentley sisHYD. It can be concluded that the WMM is an accurate model for thermal dynamics of pipeline as well as the NM. Additionally, the curve of NM is not very smooth, because the estimated time of NM $\gamma_b^{\tau} \Delta\tau + \Delta\tau/2 + (S_b^{\tau} - R_b^{\tau})/m_b^{\tau-\gamma_b^{\tau}}$ described by integer variables is discontinuous. The estimated time of WMM is continuous. Hence, the WMM may be better than the NM in the thermal dynamic simulations. The results in Fig. 3 can be found in [34].

*6.2. Small-scale System*

Fig. 4 shows the configuration of an integrated system. The EPS consists of six buses with two non-CHP units (G1 and G2), one wind farm (W1), and one back-pressure CHP unit (CHP1). Detailed data for the EPS came from [35]. The upward and downward spinning reserve requirements are both set to 60 MW. The DHS is composed of three buildings. The parameters of CHP1, the heating network parameters and the room parameters are also provided in [35]. It is assumed that there are 1,250, 250, and 1,100 rooms in the buildings of nodes 4, 5, and 6, respectively. For simplicity, we do not consider heat transfer between rooms, so the heat load and the room temperature in each room are the same. The temperature range of all rooms is set between 20 and 25°C. Electricity load and forecast wind power are shown in Fig. 5(a). The outdoor temperature curve is shown in Fig. 5(b).

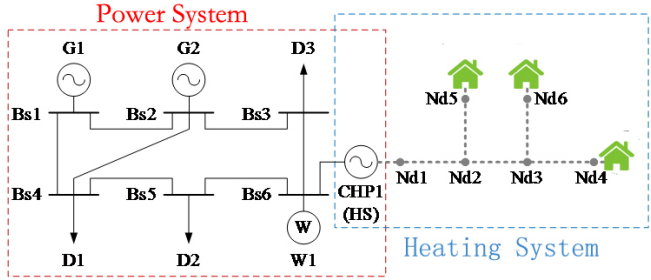

Fig. 4. Configuration of the integrated heat and electricity system.

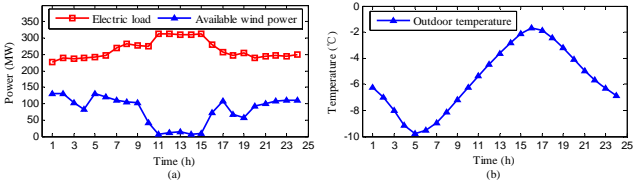

Fig. 5. (a) Electricity load and forecast wind power and (b) outdoor temperature.

The experiments are conducted in two cases.

*1) Case 1: Steady State Model*

The steady state model corresponds to the conventional dispatch model. In the steady state model, the thermal dynamics of the pipeline is not considered. The heat loss equation in the pipeline is as follows in the steady state:

$$t_{e,b}^{\tau} = t_{am,b}^{\tau} + (t_{s,b}^{\tau} - t_{am,b}^{\tau})\exp\left(-\lambda_b L/(A_b \rho c m_b^{\tau})\right) \quad (60)$$

In this case, the room temperature is fixed at 20°C. The steady-state model can be solved directly by the interior-point method with IPOPT.

The operating cost of Case 1 is $104,207. Fig. 6(a) shows the curves of outdoor temperature, indoor temperature, and heat load at Nd6. The curves are similar at Nd4 and Nd5. Heat load is defined as the total heat load in the buildings at Nd6. Compared with the lowest value of the outdoor temperature, the peak value of the heat load is delayed by about 4 hours, reflecting the thermal dynamics of the room. In Fig. 6(b), the total heat load is defined as the sum of heat loads at all nodes. Fig. 6(b) shows that the heat output and total heat load are strongly coupled in the steady state model.

*2) Case 2: IHED Model*

The IHED model is solved with the iterative algorithm proposed in Section 5 with convergence tolerance $\varepsilon = 10^{-4}$. The initial value of the mass flow is set as the solution in Case 1.

The operating cost of Case 2 is $94,018. Fig. 7(a) shows that the room temperature is not kept at the minimum when the operating cost is lowest. The indoor temperature corresponds to the heat energy stored in the room. The change in room temperature is equivalent to a change in the heat storage of the buildings. In Fig. 7(b), the heat output curve does not follow the

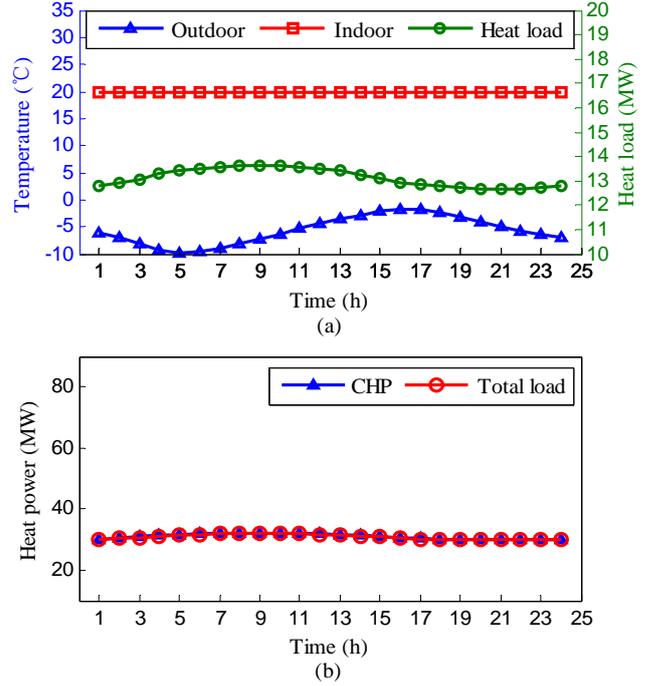

Fig. 6. (a) Outdoor/indoor temperature and heat load at Nd6 and (b) CHP heat output and total heat load in Case 1.

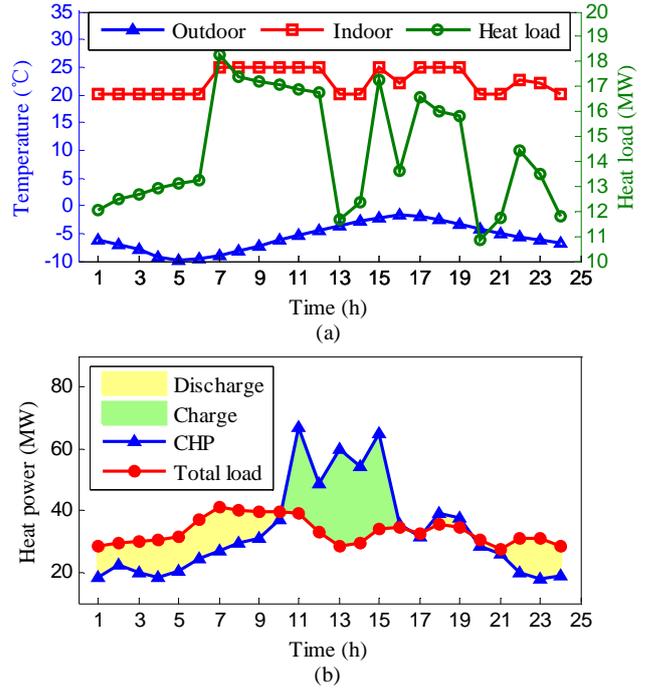

Fig. 7. (a) Outdoor/indoor temperature and heat load at Nd6 and (b) CHP heat output and total heat load in Case 2.

total heat load curve. The shaded areas in Fig. 7(b) represent the heat storage and heat release of the pipeline network. During periods 11-13, the electric load reaches the peak value and the wind power output is small; thus, CHP generation increases. Fig. 7 shows that accounting for pipeline and building thermal dynamics enhances the regulatory capability of the CHP system. Fig. 8 shows that Case 2 consumes more wind power than Case 1.

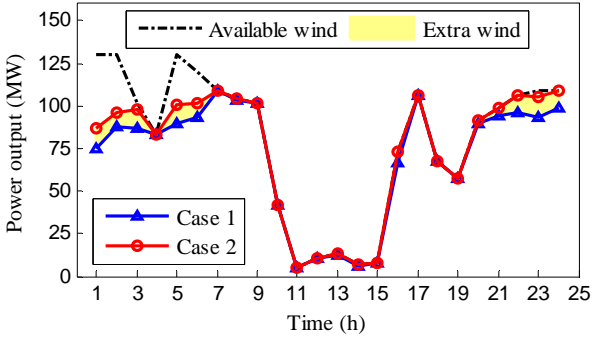

Fig. 8. Hourly wind power dispatch.

*3) Convergence Comparison*

Because the model of the buildings is linear, this IHED problem can also be solved by the iterative algorithm in [21]. We construct different scenarios by changing the forecast wind power and outdoor temperature. Then, we compare the convergence of the algorithm in this study (algorithm WMM) with that of the algorithm in [21] (algorithm NM) in these scenarios. Two stopping criteria of algorithm NM are considered as shown in (61)-(62), where the superscript $r$ is the iteration counter. The stopping criterion (61) are the same as those in [21], and (62) is a stricter stopping criterion.

$$\max_{\tau,k} | \gamma_b^{\tau,r+1} - \gamma_b^{\tau,r} | \leq 1, \max_{\tau,k} | \phi_b^{\tau,r+1} - \phi_b^{\tau,r} | \leq 1, \\ \max_{\tau,k} | K_{b,\tau,k}^{r+1} - K_{b,\tau,k}^{r} | \leq 0.1 \quad . \quad (61)$$

$$\max_{\tau,k} | \gamma_b^{\tau,r+1} - \gamma_b^{\tau,r} | < 1, \max_{\tau,k} | \phi_b^{\tau,r+1} - \phi_b^{\tau,r} | < 1, \\ \max_{\tau,k} | K_{b,\tau,k}^{r+1} - K_{b,\tau,k}^{r} | \leq 0.02 \quad . \quad (62)$$

We define scaling factors $u$ and $v$ to scale the available wind power in Fig. 5(a) and the outdoor temperature in Fig. 5(b), respectively. The convergence comparison is shown in Table 4. When algorithm NW converges with stopping criterion (61), it can be derived from (20) that the maximum error of the node temperature $t_n^\tau$ may reach 2.5°C. With stricter stopping criterion (62), the iterative process of algorithm NM is oscillating and does not converge in some scenarios. Algorithm

Table 4 Convergence Comparison

| $u$ | $v$ | Calculating time (algorithm WMM) | Calculating time (algorithm NM) with stopping criterion (61) | Calculating time (algorithm NM) with stopping criterion (62) |
|---|---|---|---|---|
| 1 | 1 | 24.53 s | 39.16 s | not convergent |
| 1 | 1.1 | 25.12 s | 41.91 s | not convergent |
| 1 | 1.15 | 26.76 s | 40.55 s | not convergent |
| 1 | 1.2 | 22.72 s | 38.48 s | 45.78 s |
| 1.1 | 1 | 24.11 s | 42.11 s | not convergent |
| 1.15 | 1 | 23.42 s | 40.98 s | 44.35 s |
| 1.2 | 1 | 25.98 s | 38.12 s | not convergent |

WMM, proposed in this paper, shows better convergence than algorithm NM.

*6.3. Large-scale Practical System*

We conduct the experiments on a large-scale system, with details in [36]. The values for the predicted electrical load and wind energy generation are based on the real data in Jilin Province of Northeast China. This large-scale system consists of a 319-bus EPS and a 40-bus DHS. The proposed sequential approach is used for solving the IHED model. Fig. 9 shows the heat output and heat load of the test system. In Fig. 9, the total heat output curve does not follow the total heat load curve, which means that the heat output and heat load are decoupled. In Fig. 10, the IHED model consumes more wind power than the steady state model.

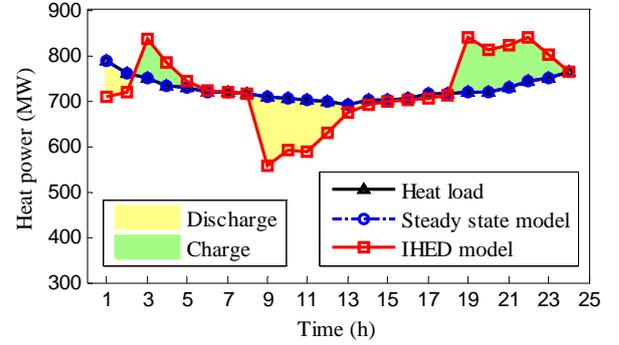

Fig. 9. CHP heat output and total heat load.

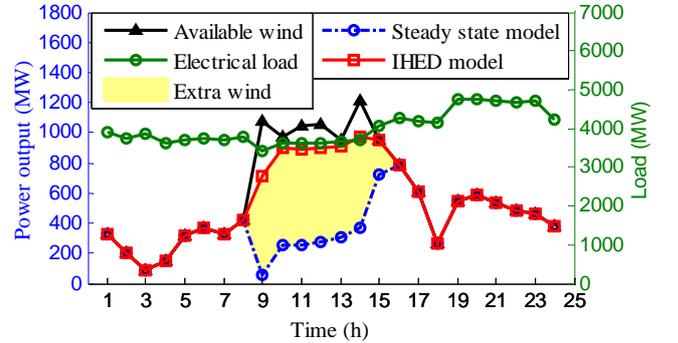

Fig. 10. Hourly wind power dispatch.

Additionally, we generate 200 scenarios by changing the forecast wind power and apply the proposed sequential approach to these scenarios. The scaling factor to scale the forecast wind power is assumed to follow a normal distribution with mean 1 and standard deviation 0.1. The Monte Carlo method is used to generate the test scenarios. The proposed algorithm WMM converges in all scenarios. And IPOPT also successfully converges to the locally optimal points in all scenarios.

## 7. Conclusions

To deal with the complexity of NM, a WMM is proposed to replace the original NM. This WMM model is an ordinary, non-linear model and eliminates 'min' and integer variables that occur in the subscripts. In the case studies, the WMM is tested and compared to the NM and Bentley sisHYD in the thermal dynamic simulations. The results of simulation show that the WMM is an accurate model for thermal dynamics of pipeline and may be better than the NM in the thermal dynamic simulations. After applying the WMM model, we present an IHED model considering the thermal dynamics of pipelines and buildings to increase dispatch flexibility. Then, based on the GBD, we describe an iterative algorithm to solve the IHED model. Besides the iterative algorithm, a sequential approach combined with the iterative algorithm and IPOPT is proposed to improve the optimality of the solution. A small-scale system and a large-scale practical system are used to demonstrate the convergence and effectiveness of the proposed solution strategy. The simulation results for the test systems indicate that the proposed model can provide more operating flexibility and decrease wind curtailment in the integrated heat and electricity system. Convergence of the algorithm WMM is also validated by numerical examples, compared with algorithm NM.

There are several issues that need further study. The WMM can also be used in these contexts. First, one problem is how to measure the energy storage capacity available for each dispatch period in a pipeline network. Moreover, based on this IHED model, we can further study the demand response in an integrated heat and electricity system. Finally, how to analyze the uncertainty of renewable energy and loads remains as an important topic.

**Acknowledgements**

This work was supported in part by the National Key R&D Program of China (2016YFB0901300), in part by the National Natural Science Foundation of China (NSFC) (51537006), in part by the China Postdoctoral Science Foundation under Grants 2016M600091 and 2017T100078.

**References**

[1] Meibom P, Hilger KB, Madsen H, Vinther D. Energy Comes Together in Denmark: The Key to a Future Fossil-Free Danish Power System. IEEE Power & Energy Magazine. 2013;11:46-55.
[2] Zaman K, Moemen AE. Energy consumption, carbon dioxide emissions and economic development: Evaluating alternative and plausible environmental hypothesis for sustainable growth. Renewable & Sustainable Energy Reviews. 2017;74:1119-30.
[3] Xu SC, He ZX, Long RY. Factors that influence carbon emissions due to energy consumption in China: Decomposition analysis using LMDI. Applied Energy. 2014;127:182-93.
[4] Meibom P, Hilger KB, Madsen H, Vinther D. Energy Comes Together in Denmark: The Key to a Future Fossil-Free Danish Power System. IEEE Power & Energy Magazine. 2013;11:46-55.
[5] Kienzle F, Favre-Perrod P, Arnold M, Andersson G. Multi-energy delivery infrastructures for the future. First International Conference on Infrastructure Systems and Services: Building Networks for a Brighter Future, 2008. p. 1-5.
[6] Omalley M, Kroposki B. Energy Comes Together: The Integration of All Systems [Guest Editorial]. IEEE Power & Energy Magazine. 2013;11:18-23.
[7] Krause T, Andersson G, Frohlich K, Vaccaro A. Multiple-Energy Carriers: Modeling of Production, Delivery, and Consumption. Proceedings of the IEEE. 2010;99:15-27.
[8] Wu J, Yan J, Jia H, Hatziargyriou N, Djilali N, Sun H. Integrated Energy Systems. Applied Energy. 2016;167:155-7.
[9] Mancarella P. MES (multi-energy systems): An overview of concepts and evaluation models. Energy. 2014;65:1-17.
[10] Li J et al., 2014 China Wind Power Review and Outlook. Beijing, China: Chinese Renewable Energy Ind. Assoc. (CREIA), 2014.
[11] Chen X, Kang C, O'Malley M, Xia Q, Bai J, Liu C, et al. Increasing the Flexibility of Combined Heat and Power for Wind Power Integration in China: Modeling and Implications. IEEE Transactions on Power Systems. 2015;30:1848-57.
[12] Yang Y, Wu K, Yan X, Gao J, Long H. The large-scale wind power integration using the integrated heating load and heating storage control. PowerTech, 2015 IEEE Eindhoven2015.
[13] Bahmani-Firouzi B, Farjah E, Seifi A. A new algorithm for combined heat and power dynamic economic dispatch considering valve-point effects. Energy. 2013;52:320-32.
[14] Ghorbani N. Combined heat and power economic dispatch using exchange market algorithm. International Journal of Electrical Power & Energy Systems. 2016;82:58-66.
[15] Roy PK, Paul C, Sultana S. Oppositional teaching learning based optimization approach for combined heat and power dispatch. Electrical Power & Energy Systems. 2014;57:392-403.
[16] Li J, Fang J, Zeng Q, Chen Z. Optimal operation of the integrated electrical and heating systems to accommodate the intermittent renewable sources. Applied Energy. 2016;167:244-54.
[17] Awad N, Wu J, Ekanayake J, Jenkins N. Integrated operation of an energy MicroGrid with islanded electricity network, 21st International Conference on Electricity Distribution, Frankfurt, Germany, 2011.
[18] Gu, Wei, et al. "Optimal operation for integrated energy system considering thermal inertia of district heating network and buildings." Applied Energy. 2017:199:234-246.
[19] Lin C, Wu W, Zhang B, Sun Y. Decentralized Solution for Combined Heat and Power Dispatch through Benders Decomposition. IEEE Transactions on Sustainable Energy. 2017.
[20] Pan Z, Guo Q, Sun H. Feasible region method based integrated heat and electricity dispatch considering building thermal inertia. Applied Energy. 2017;192.
[21] Li Z, Wu W, Shahidehpour M, Wang J, Zhang B. Combined Heat and Power Dispatch Considering Pipeline Energy Storage of District Heating Network. IEEE Transactions on Sustainable Energy. 2015;7:12-22.
[22] IPOPT Homepage. Available: https://projects.coin-or.org/Ipopt
[23] Lahdelma R, Hakonen H. An efficient linear programming algorithm for combined heat and power production. European Journal of Operational Research. 2003;148:141-51.
[24] Guo T, Henwood MI, Van Ooijen M. An algorithm for combined heat and power economic dispatch. Power Systems IEEE Transactions on. 1996;11:1778-84.
[25] Sinha N, Saikia LC, Malakar T. Optimal solution for non-convex combined heat and power dispatch problems using Differential Evolution. IEEE International Conference on Computational Intelligence and Computing Research2010. p. 1-5.
[26] Liu X. Combined heat and power dispatch with wind power: A stochastic model and solutions. Power and Energy Society General Meeting, IEEE2010, 2010. p. 1-6.
[27] Stephenson DG, Mitalas GP. Cooling load calculations by thermal response factor method. ASHRAE Trans. 1967;73.
[28] Benonysson A, Dynamic modelling and operational optimization of district heating systems, Ph.D. dissertation, Lab. Heating Air Conditioning, Tech. Univ., Lyngby, Denmark, 1991.
[29] Zhao H, Analysis, modelling and operational optimization of district heating systems, Ph.D. dissertation, Centre District Heating Technol., Tech. Univ. Denmark, Lyngby, Denmark, 1995.
[30] Geoffrion AM. Generalized Benders decomposition. Journal of Optimization Theory and Applications. 1972;10:237-60.
[31] Grossmann IE. MINLP: Outer Approximation Algorithm. In: Floudas CA, Pardalos PM, editors. Encyclopedia of Optimization. Boston, MA: Springer US; 2001. p. 1459-63.
[32] CPLEX Homepage. Available: http://www-01.ibm.com/software/commerce/optimization/cplex-optimizer
[33] Bentley sisHYD. Available: https://www.bentley.com/
[34] Data and results of thermal dynamic simulations. Available: https://drive.google.com/open?id=1eFi_oBEJDYRuIymzxH5GM6jxD_AbHWl8
[35] Test data for small-scale system. Available: https://drive.google.com/open?id=14VCNC4E1fOKImTIuzzF29955SqP7Xawx
[36] Test data for large-scale system. Available: https://drive.google.com/open?id=1ACtIXkjdcYJ9kj-YPxqba-AYUtTJwqB0